\documentclass[aps,preprint,showpacs,preprintnumbers,amsmath,amssymb]{revtex4}
\usepackage{amsmath,mathrsfs,amsbsy,color,graphicx,bm,amsthm,amsfonts}
\usepackage{units}
\usepackage{bbm}
\usepackage{times}
\usepackage{dcolumn}
\usepackage{mathrsfs}
\usepackage{amsmath,amssymb,epsfig}
\usepackage{amsmath}
\newcommand{\udots}{\mathinner{\mskip1mu\raise1pt\vbox{\kern7pt\hbox{.}}
\mskip2mu\raise4pt\hbox{.}\mskip2mu\raise7pt\hbox{.}\mskip1mu}}
\begin{document}
\title{Does the survival and sudden death of quadripartite steering in curved spacetime truly depend on multi-directionality?}
\author{Xiaobao Liu$^1$\footnote{xiaobaoliu@hotmail.com}, Wentao Liu$^{2}$\footnote{wentaoliu@hunnu.edu.cn}, Si-Han Shang$^3$, Shu-Min Wu$^3$\footnote{smwu@lnnu.edu.cn (corresponding author)}}
\affiliation{$^1$ Department of physics and electrical engineering, Liupanshui Normal University, Liupanshui
 553004, Guizhou, China \\
 $^2$ Lanzhou Center for Theoretical Physics, Key Laboratory of Theoretical  \\ Physics of Gansu Province, Key Laboratory of Quantum Theory and Applications  of MoE, Gansu Provincial Research Center for Basic Disciplines of Quantum Physics, Lanzhou University, Lanzhou 730000, China\\
$^3$
Department of Physics, Liaoning Normal University, Dalian 116029, China
 }


\begin{abstract}
We systematically investigate the directional dependence of Gaussian quadripartite quantum steering and its redistribution among different modes in the background of a Schwarzschild black hole. For physically accessible sectors, we identify three distinct behaviors: (i) steering from non-gravitational to gravitational observers undergoes sudden death at maximal asymmetry with the Hawking temperature, marking the crossover from two-way to one-way steerability; (ii) steering in the opposite direction decays monotonically and vanishes only in the extreme black hole limit, highlighting its directional sensitivity to spacetime curvature;  (iii)  steering from hybrid gravitational-non-gravitational partitions to non-gravitational mode persists at a finite asymptotic value set by the initial squeezing parameter. Moreover, all inaccessible steerings generated by the Hawking effect exhibit an intrinsic asymmetry, with their specific behavior being strongly dependent on the steering direction.
\end{abstract}

\vspace*{0.5cm}
\maketitle
\section{introduction}
Quantum steering, first introduced by Schr\"{o}dinger in response to the EPR paradox in 1936 \cite{Q1,Q2}, describes the ability of one party to asymmetrically influence the state of a distant quantum system through shared entanglement \cite{Q3,Q4,Q6,Q7}. This unique feature makes steering particularly valuable for device-independent entanglement verification. The concept was rigorously formalized in 2007 by Wiseman et al. \cite{Q8}, who established an operational framework that positions steering as an intermediate form of quantum correlation stronger than entanglement but generally weaker than Bell nonlocality. A central aspect of this framework is the intrinsic directionality of steering \cite{Q9,Q10,Q12}, which has been experimentally demonstrated in both continuous-variable and discrete-variable systems \cite{Q13,Q14,Q15,Q16}. Extending bipartite steering to multipartite systems introduces richer and more complex correlation structures. In tripartite configurations, for instance, Reid's monogamy relation imposes constraints: a single party can steer two others simultaneously, but two independent parties cannot both steer the same target \cite{Q17,ZAVM1}. Building on these insights, Teh et al. have derived criteria for multipartite steering in continuous-variable systems, demonstrated methods to generate and detect genuine N-partite steerable states at network nodes, and analyzed both the distribution and monogamy properties of steering \cite{Z0924}. Furthermore, multipartite systems exhibit hierarchical steering patterns, encompassing genuine multipartite, reduced bipartite, and collective steering regimes. These intricate properties make multipartite steering a powerful tool for designing and optimizing quantum networks, and they underpin novel protocols for scalable quantum communication, establishing steering as a practical and versatile quantum resource \cite{Q5}.

Within the framework of general relativity, the gravitational collapse of sufficiently massive stars inevitably results in the formation of black holes. Landmark observations, including the first black hole image and the detection of gravitational waves, have profoundly advanced our understanding of black hole formation, dynamical evolution, and the physics of strong gravitational fields \cite{L16,L17,L18,L19,L20,L21,L22}. Beyond astrophysics, these discoveries also shed light on the fundamental structure of spacetime itself.
The causal distinction between the interior and exterior of a black hole's event horizon gives rise to a host of exotic phenomena. Among the most striking is Hawking radiation, which originates from quantum field fluctuations near the horizon through the creation and annihilation of virtual particle-antiparticle pairs \cite{L23,L24,L25}. Serving as a unique bridge between quantum theory and gravity, Hawking radiation is central to the black hole information paradox, a long standing challenge in theoretical physics \cite{L26,L27,L28}. The influence of Hawking radiation, together with its flat spacetime analogue, the Unruh effect, on quantum resources and information theoretic tasks has been widely studied \cite{zgkl1,zgkl2,zgkl3,zgkl4,zgkl5,zgkl6,zgkl7,zgkl8,TH1,TH2,TH3,BRQ1,BRQ2,BRQ3,BRQ4,BRQ5,BRQ6,BRQ7,BRQ8,BRQ9,BRQ10,BRQ11,
BRQ12,BRQ13,BRQ14,BRQ15,BRQ16,BRQ17,BRQ18,BRQ19,BRQ20,BRQ21,BRQ22,BRQ23,BRQ24,
QQQ1,QET1,QQBJ2,QQBJ3,TRQ1,TRQ2,TRQ3,TRQ4,TRQ5,TRQ6,TRQ7,TRQ8,TRQ9,TRQ10,TRQ11,
aqe1,aqe2,ST1,ST2,ST3,ST4,QBP1,QBP2,QBP3,QBP4,QFK1,QFK2}. Within the relativistic framework, however, multipartite entanglement and coherence are generally regarded as direction-independent quantum resources \cite{TRQ1,TRQ2,TRQ3,TRQ4,TRQ5,TRQ6,TRQ7,TRQ8,TRQ9,TRQ10,TRQ11,aqe1,aqe2}.
It is well known that, unlike entanglement and coherence, multipartite quantum steering exhibits direction-dependent properties. Yet, in the context of the Schwarzschild black hole, these directional features remain unclear. Notably, in certain directions, multipartite quantum steering may possess the ability to counteract the Hawking effect of the black hole, thereby offering advantages for implementing relativistic quantum information processing tasks.

In this work, we perform a detailed analysis of Gaussian quadripartite steering in the background of a Schwarzschild black hole. We consider four modes: $A$, $B$, and $C$, associated with Kruskal observers Alice, Bob, and Charlie; $D$, corresponding to a Schwarzschild observer David outside the event horizon; and $\bar{D}$, observed by anti-David inside the horizon. Employing the Bogoliubov transformation between the Kruskal and Schwarzschild vacua, we derive the phase-space evolution of an initially Gaussian four-mode state under Hawking radiation. Our theoretical calculations yield analytical expressions for both physically accessible and inaccessible four-mode steering, providing a rigorous quantitative framework to probe quantum correlations in curved spacetime. For physically accessible modes, three prominent behaviors are observed: (i) steering from non-gravitational to gravitational modes experiences sudden death at maximal asymmetry with the Hawking temperature, signaling the crossover from two-way to one-way steerability; (ii) steering in the opposite direction decays monotonically  and vanishes only in the extremal limit, reflecting its directional sensitivity to Hawking effect; (iii) hybrid steering, where gravitational and non-gravitational modes jointly influence a non-gravitational mode, saturates at a finite asymptotic value determined by the initial squeezing. For inaccessible modes, the Hawking effect induces steering that can exhibit sudden birth and pronounced asymmetry, demonstrating that Hawking radiation not only degrades but also redistributes multipartite steerability. These findings illuminate the intricate interplay between black hole physics and the directional structure of quantum correlations.

The paper is structured as follows. In Sec. II, we briefly review the definition and quantification of Gaussian quantum steering. Sec. III presents a description of the Hawking effect in terms of an effective Gaussian channel. In Sec. IV, we investigate the distribution of Gaussian quadripartite steering in the Schwarzschild black hole background. Finally, Sec. V summarizes our main findings and conclusions.

\section{ Measure of Gaussian quantum steering  \label{GSCDGE}}
In this section, we briefly review the concept of Gaussian bipartite steering. Consider a bipartite Gaussian state consisting of
$(n_{A} + m_{B})$ modes, where the quadratures of each mode position (amplitude) and momentum (phase) are collected into a column vector $\hat{\xi}:=(\hat{x}^{A}_{1}, \hat{p}^{A}_{1},..., \hat{x}^{A}_{n_A}, \hat{p}^{A}_{n_A}, \hat{x}^{B}_{1}, \hat{p}^{B}_{1},..., \hat{x}^{B}_{m_B}, \hat{p}^{B}_{m_B})^{T}$, satisfying the canonical commutation relations $[\hat{x}_{j}, \hat{p}_{k}]=2i\delta_{jk}$.
The state is fully characterized by its covariance matrix ${\sigma_{AB}}$ with elements $$(\sigma_{AB})_{ij}=\langle\hat{\xi}_{i}\hat{\xi}_{j}+\hat{\xi}_{j}\hat{\xi}_{i}\rangle/2-\langle\hat{\xi}_{i}\rangle\langle\hat{\xi}_{j}\rangle,$$ which can be expressed in block form as
\begin{eqnarray}\label{zs1}
\sigma_{AB}= \left(
\begin{array}{cccc}
A\qquad&C&\\
C^{T}\qquad&B&\\
 \end{array}
 \right),
\end{eqnarray}
where $A$ and $B$ correspond to the reduced covariance matrices of subsystems $A$ and $B$, and $C$ captures the correlations between them.

The steerability from Alice to Bob under Gaussian measurements is quantified by \cite{L85}
\begin{eqnarray}\label{w3}
\mathcal{G}^{A\rightarrow B}(\sigma_{AB}):=\max\left\{0,-\Sigma_{j:\bar{\nu}_{j}^{AB\backslash A}<1}\ln(\bar{\nu}_{j}^{AB\backslash A})\right\},
\end{eqnarray}
where $\bar{\nu}_{j}^{AB\backslash A}(j=1,..., m_{B})$ are the symplectic eigenvalues of the Schur complement of $A$ defined as $\bar{\sigma}_{AB\backslash A}=B-C^{T}A^{-1}C$.
This measure is monotonic under Gaussian local operations and classical communication and vanishes if Alice cannot steer Bob \cite{L86}. Experimentally, it has been verified via covariance matrix reconstruction using Gaussian cluster states \cite{L87}. For a specific Gaussian state  $\sigma_{AB}$, the steerability simplifies to
\begin{eqnarray}\label{wA1}
\mathcal{G}^{A\rightarrow B}(\sigma_{AB})&=\max\left\{0,\frac{1}{2}\ln\frac{\det A}{\det \sigma_{AB}}\right\} =&\max\{0,S(A)-S(\sigma_{AB})\},
\end{eqnarray}
where $S{(\sigma)}=\frac{1}{2}\ln(\det \sigma)$ is the R\'{e}nyi-2 entropy \cite{L88}.
By exchanging the roles of $A$ and $B$, one obtains the steerability in the opposite direction
\begin{eqnarray}\label{wA2}
\mathcal{G}^{B\rightarrow A}(\sigma_{AB})&=\max\left\{0,\frac{1}{2}\ln\frac{\det B}{\det \sigma_{AB}}\right\} =&\max\{0,S(B)-S(\sigma_{AB})\}.
\end{eqnarray}
This formalism provides a convenient way to quantify the directional correlations inherent in Gaussian states, allowing one to determine how strongly one subsystem can influence the other under Gaussian measurements.

\section{Hawking effect in terms of effective Gaussian channel } \label{GSCDGE}
The Schwarzschild black hole metric \cite{Z1} is given by
\begin{eqnarray}\label{l1}
ds^{2}=-(1-\frac{2M}{r})dt^{2}+(1-\frac{2M}{r})^{-1}dr^{2}+r^{2}(d\theta^{2}+\sin^{2}\theta d\varphi^{2}),
\end{eqnarray}
where $M$ represents the black hole mass. The dynamics of a massless scalar field in this spacetime are governed by the Klein-Gordon equation \cite{Z11}
\begin{eqnarray}\label{S2}
\frac{1}{\sqrt{-g}}\frac{\partial}{\partial x^{\mu}}(\sqrt{-g}g^{\mu\nu}\frac{\partial\Psi}{\partial x^{\nu}})=0.
\end{eqnarray}
Assuming a separable solution, the field can be decomposed as
\begin{eqnarray}\label{S3}
\Psi_{\omega lm}=\frac{1}{R(r)}\chi_{\omega l}(r)Y_{lm}(\theta,\varphi)e^{-i\omega t},
\end{eqnarray}
where $Y_{lm}(\theta,\varphi)$ denotes the spherical harmonic on the unit two-sphere. The corresponding radial equation  satisfies
\begin{eqnarray}\label{S4}
\frac{d^{2}\chi_{\omega l}}{dr^{2}_{\ast}}+[\omega^{2}-V(r)]\chi_{\omega l}=0,
\end{eqnarray}
with the potential $V(r)$ defined as
\begin{eqnarray}\label{z5}
V(r)=\frac{\sqrt{f(r)h(r)}}{R(r)}\frac{d}{dr}\left[\sqrt{f(r)h(r)}\frac{dR(r)}{dr}\right]+\frac{l(l+1)f(r)}{R^{2}(r)}.
\end{eqnarray}
Solving Eq.(\ref{S4}) yields the incoming mode
\begin{eqnarray}\label{S5}
\Psi_{in,\omega lm}=e^{-i\omega \nu}Y_{lm}(\theta,\varphi),
\end{eqnarray}
which is analytic throughout the spacetime, where $\nu=t+r_{\ast}$. The outgoing modes are defined differently inside and outside the event horizon
\begin{eqnarray}\label{q6}
\Psi_{out,\omega lm}(r>r_{+})=e^{-i\omega \mu}Y_{lm}(\theta,\varphi),
\end{eqnarray}
\begin{eqnarray}\label{q7}
\Psi_{out,\omega lm}(r<r_{+})=e^{i\omega \mu}Y_{lm}(\theta,\varphi),
\end{eqnarray}
with $\mu=t-r_{\ast}$. These solutions form orthogonal sets in their respective regions. For quantization in the black hole exterior, the scalar field can be expanded as
\begin{eqnarray}\label{q8}
\Phi_{out}&=&\sum_{lm}\int d\omega[b_{in,\omega lm}\Psi_{out,\omega lm}(r<r_{+})
+b_{in,\omega lm}^{\dag}\Psi_{out,\omega lm}^{\ast}(r<r_{+})\notag\\
&+&b_{out,\omega lm}\Psi_{out,\omega lm}(r>r_{+})+b_{out,\omega lm}^{\dag}\Psi_{out,\omega lm}^{\ast}(r>r_{+})],
\end{eqnarray}
where $b_{in,\omega lm}$ ($b_{in,\omega lm}^{\dagger}$) and $b_{out,\omega lm}$ ($b_{out,\omega lm}^{\dagger}$) are the annihilation (creation) operators associated with the interior and exterior regions, respectively \cite{BRQ11}. The corresponding  Schwarzschild vacuum states satisfy $b_{in,\omega lm}|0\rangle_{in}=b_{out,\omega lm}|0\rangle_{out}=0.$

The Schwarzschild spacetime can be expressed using lightlike Kruskal coordinates defined by
\begin{eqnarray}\label{Z1}
U&=&4Me^{-\frac{\mu}{4M}},V=4Me^{\frac{\nu}{4M}},\quad \mathrm {if}\ r<r_{+};\notag\\
U&=&-4Me^{-\frac{\mu}{4M}},V=4Me^{\frac{\nu}{4M}},\ \mathrm {if}\ r>r_{+}.
\end{eqnarray}
In these coordinates, the outgoing Schwarzschild modes take the form
\begin{eqnarray}\label{Z2}
\Psi_{out,\omega lm}(r<r_{+})=e^{-4i\omega M\ln[-\frac{U}{4M}]}Y_{lm}(\theta,\varphi),
\end{eqnarray}
\begin{eqnarray}\label{Z3}
\Psi_{out,\omega lm}(r>r_{+})=e^{4i\omega M\ln[\frac{U}{4M}]}Y_{lm}(\theta,\varphi).
\end{eqnarray}
Following the approach of Ruffini and Damour \cite{Z3}, one can construct the outgoing modes that are analytic across the  event horizon in Kruskal coordinates
\begin{eqnarray}\label{S10}
\Psi_{I,\omega lm}=e^{(\pi\omega/2\kappa)}\Psi_{out,\omega lm}(r>r_{+})
+e^{-(\pi\omega/2\kappa)}\Psi_{out,\omega lm}^{\ast}(r<r_{+}),
\end{eqnarray}
\begin{eqnarray}\label{S11}
\Psi_{II,\omega lm}=e^{(-\pi\omega/2\kappa)}\Psi_{out,\omega lm}^{\ast}(r>r_{+})
+e^{(\pi\omega/2\kappa)}\Psi_{out,\omega lm}(r<r_{+}).
\end{eqnarray}
Thus,  quantization of the scalar field in terms of these Kruskal modes leads to
\begin{eqnarray}\label{S12}
\Phi_{out}&=&\sum_{lm} \int d\omega[2\sinh{(\pi\omega/\kappa)]^{-1/2}}[a_{out,\omega lm}\Psi_{I,\omega lm}\notag\\
&+&a_{out,\omega lm}^{\dagger}\Psi_{I,\omega lm}^{\ast}+a_{in,\omega lm}\Psi_{II,\omega lm}+a_{in,\omega lm}^{\dagger}\Psi_{II,\omega lm}^{\ast}].
\end{eqnarray}
Here, the annihilation operator $a_{out,\omega lm}$ defines the Kruskal vacuum $a_{out,\omega lm}|0\rangle_{K}=0.$ Comparing Eqs.(\ref{q8}) and (\ref{S12}) allows one to derive the Bogoliubov transformations connecting Schwarzschild and Kruskal operators
\begin{eqnarray}\label{S14}
a_{out,\omega lm}=\frac{b_{out,\omega lm}}{\sqrt{1-e^{-\omega/T}}}-\frac{b_{in,\omega lm}^{\dag}}{\sqrt{e^{\omega/T}-1}},
\end{eqnarray}
\begin{eqnarray}\label{S15}
a_{out,\omega lm}^{\dag}=\frac{b_{out,\omega lm}^{\dag}}{\sqrt{1-e^{-\omega/T}}}
-\frac{b_{in,\omega lm}}{\sqrt{e^{\omega/T}-1}},
\end{eqnarray}
where $T=\frac{1}{8\pi M}$ denotes the Hawking temperature \cite{BRQ11}.

After proper normalization, the Kruskal vacuum state in the Schwarzschild black hole can be represented as a two-mode squeezed state that is maximally entangled between the exterior and interior regions
 \begin{eqnarray}\label{S23}
|0_{\omega}\rangle_{K}=\frac{1}{\cosh s}\underset{n=0}{\overset{\infty}{\sum}}\tanh^{n}s|n_{\omega}\rangle_{out}|n_{\omega}\rangle_{in}=\hat{U}|0\rangle_{out}|0\rangle_{in},
\end{eqnarray}
where $s$ is related to the Hawking temperature and satisfies $\cosh s=\frac{1}{\sqrt{1-e^{-\omega/T}}}$. The operator $$\hat{U}=\exp[s(b^{\dag}_{out}b^{\dag}_{in}-b_{out}b_{in})]$$
defines a two-mode squeezing transformation acting on the vacuum states of the exterior and interior modes. Physically, $\hat{U}$  generates pairs of correlated particles across the event horizon, encoding the entanglement structure that underlies Hawking radiation.
Importantly, $\hat{U}$ is a Gaussian unitary operation, which preserves the Gaussian character of the quantum states. Its action can be conveniently described in the symplectic phase-space representation, where the corresponding two-mode squeezing matrix takes the form
\begin{eqnarray}\label{S25}
S_{D,\bar{D}}(s)= \left(
\begin{array}{cccc}
\cosh(s)&0& \sinh(s)&0\\
0&\cosh(s)&0&-\sinh(s)\\
\sinh(s)&0&\cosh(s)&0\\
0&-\sinh(s)&0&-\cosh(s)
 \end{array}
 \right).
\end{eqnarray}

\section{Distribution of Gaussian quadripartite steering in curved spacetime \label{GSCDGE}}
In an asymptotically flat spacetime, the four-mode squeezed vacuum Gaussian state shared among Alice, Bob, Charlie, and David is described by the covariance matrix \cite{Z6,Z21}
\begin{eqnarray}\label{S26}
\sigma_{ABCD}(r)=\left(\begin{array}{cccc}
\cosh^2(r)I_2&\frac{1}{2}\sinh(2r)\sigma_z&\sinh^2(r)I_2&\frac{1}{2}\sinh(2r)\sigma_z\\
\frac{1}{2}\sinh(2r)\sigma_z&\cosh^2(r)I_2&\frac{1}{2}\sinh(2r)\sigma_z&\sinh^2(r)I_2\\
\sinh^2(r)I_2&\frac{1}{2}\sinh(2r)\sigma_z&\cosh^2(r)I_2&\frac{1}{2}\sinh(2r)\sigma_z\\
\frac{1}{2}\sinh(2r)\sigma_z&\sinh^2(r)I_2&\frac{1}{2}\sinh(2r)\sigma_z&\cosh^2(r)I_2\\
\end{array}\right),
\end{eqnarray}
where $r$ is initial squeezing parameter,  $I_2$ denotes the $2\times2$ identity matrix, and $\sigma_z$ is the Pauli matrix along the $z$ direction. In this configuration, Alice, Bob, and Charlie remain static in the asymptotically flat region, while David approaches the event horizon of the black hole. The transition from Kruskal to Schwarzschild modes induces a symplectic map $S_{D\bar{D}}$ in the phase space, which splits David's original mode $D$ into two parts: the exterior mode $D$ (outside the event horizon) and the interior mode $\bar D$ (inside the event horizon). Consequently, the complete Gaussian system must be described as a five-mode state: Alice, Bob, and Charlie correspond to modes $A$, $B$, and $C$ observed by Kruskal observers, David corresponds to Schwarzschild mode $D$, and $\bar D$ is associated with the so-called ``anti-David" observer beyond the event horizon. The covariance matrix of this extended state reads \cite{BRQ7}
\begin{eqnarray}\label{S27}
\nonumber\sigma_{ABCD\bar D}(r,s) &=& \big[I_{ABC}\oplus  S_{D, \bar D}(s)\big] \big[\sigma_{ABCD}(r)\oplus I_{\bar D}\big]\\&& \big[I_{ABC}\oplus  S_{D, \bar D}(s)\big]^T\,\nonumber\\
 &=& \left(
\begin{array}{ccccc}
\mathcal{\sigma}_{A}(r) & \mathcal{E}_{AB}(r) & \mathcal{E}_{AC}(r) & \mathcal{E}_{AD}(r,s)&\mathcal{E}_{A\bar D}(r,s) \\
\mathcal{E}^{\sf T}_{AB}(r) & \mathcal{\sigma}_{B}(r) & \mathcal{E}_{BC}(r)& \mathcal{E}_{BD}(r,s) & \mathcal{E}_{B\bar D}(r,s) \\
\mathcal{E}^{\sf T}_{AC}(r)  & \mathcal{E}^{\sf T}_{BC}(r)&  \mathcal{\sigma}_{C}(r) & \mathcal{E}_{CD}(r,s) & \mathcal{E}_{C\bar D}(r,s) \\
\mathcal{E}^{\sf T}_{AD}(r,s) & \mathcal{E}^{\sf T}_{BD}(r,s)  & \mathcal{E}_{CD}^{\sf T}(r,s)& \mathcal{\sigma}_{D}(r,s)& \mathcal{E}_{D\bar D}(r,s)\\
\mathcal{E}^{\sf T}_{A\bar D}(r,s) & \mathcal{E}^{\sf T}_{B\bar D}(r,s) & \mathcal{E}^{\sf T}_{C\bar D}(r,s)  & \mathcal{E}^{\sf T}_{D\bar D}(r,s)&  \mathcal{\sigma}_{\bar D}(r,s)
\end{array}
\right)\,.
\end{eqnarray}

In Eq.(\ref{S27}), the block $\big[\sigma_{ABCD}(r)\oplus I_{\bar D}\big]$ represents the initial covariance matrix of the system. The diagonal entries are given by
\begin{gather}
\mathcal{\sigma}_{A}(r)=\mathcal{\sigma}_{B}(r)=\mathcal{\sigma}_{C}(r)=\cosh^2(r)I_2,\nonumber\\ \mathcal{\sigma}_{D}(r,s)=[\cosh^2(r)\cosh^2(s)+\sinh^2(s)]I_2,\nonumber\\ \nonumber\mathcal{\sigma}_{\bar D}(r,s)=[\cosh^2(s)+\cosh^2(r)\sinh^2(s)]I_2.\nonumber
\end{gather}
The off-diagonal blocks take the following forms
\begin{gather}\nonumber\mathcal{E}_{AB}(r)=\mathcal{E}_{BC}(r)=\left(\begin{array}{cccc}
\frac{1}{2}\sinh(2r)\qquad 0\\
0 \qquad -\frac{1}{2}\sinh(2r)\\
\end{array}\right),\end{gather}
\begin{gather}\mathcal{E}_{AC}(r)=\left(\begin{array}{cccc}
\sinh^2(r)\qquad 0\\
0\qquad \sinh^2(r)\\
\end{array}\right),\end{gather}
\begin{gather}\mathcal{E}_{AD}(r,s)=\mathcal{E}_{CD}(r,s)=\left(\begin{array}{cccc}
\frac{1}{2}\cosh(s)\sinh(2r)\qquad 0\\
0\qquad -\frac{1}{2}\cosh(s)\sinh(2r)\\
\end{array}\right),\end{gather}
\begin{gather}\mathcal{E}_{BD}(r,s)=\left(\begin{array}{cccc}
\cosh(s)\sinh^2(r)\qquad 0\\
0\qquad \cosh(s)\sinh^2(r)\\
\end{array}\right),\end{gather}
\begin{gather}\mathcal{E}_{A\bar D}(r,s)=\mathcal{E}_{C\bar D}(r,s)=\left(\begin{array}{cccc}
\frac{1}{2}\sinh(2r)\sinh(s)\qquad 0\\
0\qquad -\frac{1}{2}\sinh(2r)\sinh(s)\\
\end{array}\right),\end{gather}
\begin{gather}\nonumber\mathcal{E}_{B\bar D}(r,s)=\left(\begin{array}{cccc}
\sinh^2(r)\sinh(s)\qquad 0\\
0\qquad -\sinh^2(r)\sinh(s)\\
\end{array}\right),\end{gather} and \begin{gather}\nonumber\mathcal{E}_{D\bar D}(r,s)=\left(\begin{array}{cccc}
\cosh(s)\sinh(s)+\cosh^2(r)\cosh(s)\sinh(s)\qquad 0\\
0\qquad -\cosh(s)\sinh(s)-\cosh^2(r)\cosh(s)\sinh(s)\\
\end{array}\right).\nonumber \end{gather}

\subsection{Gaussian quadripartite steering between initially correlated modes}
Since the black hole interior is causally inaccessible to external observers, the parties Alice, Bob, Charlie, and David have no access to mode $\bar D$. By tracing out this inaccessible mode, we obtain the reduced covariance matrix for the four observable modes $A$, $B$, $C$, and $D$ as
\begin{eqnarray}\label{B2}
\sigma_{ABCD}(r,s) &=&
\left(\begin{array}{ccccc}
\mathcal{\sigma}_{A}(r) & \mathcal{E}_{AB}(r) & \mathcal{E}_{AC}(r) & \mathcal{E}_{AD}(r,s)  \\
\mathcal{E}^{\sf T}_{AB}(r) &  \mathcal{\sigma}_{B}(r) &\mathcal{E}_{BC}(r) & \mathcal{E}_{BD}(r,s) &  \\
\mathcal{E}^{\sf T}_{AC}(r)  &\mathcal{E}^{\sf T}_{BC}(r) &  \mathcal{\sigma}_{C}(r) & \mathcal{E}_{CD}(r,s) &  \\
\mathcal{E}^{\sf T}_{AD}(r,s) & \mathcal{E}^{\sf T}_{BD}(r,s) & \mathcal{E}^{\sf T}_{CD}(r,s) & \mathcal{\sigma}_{D}(r,s) & \\
\end{array}
\right)\,.
\end{eqnarray}
Employing Eqs.(\ref{wA1}) and (\ref{wA2}), the closed-form expressions of Gaussian quadripartite steering can be derived for different $1\rightarrow 3$ and $3\rightarrow 1$ steering configurations associated with the state in Eq.(\ref{B2})
\begin{eqnarray}\label{S31}
\mathcal{G}^{A \rightarrow\ BCD}(\sigma_{ABCD})&=&\mathcal{G}^{C \rightarrow\ ABD}(\sigma_{ABCD})=\mathcal{G}^{ABC \rightarrow\ D}(\sigma_{ABCD}) \\ \nonumber
&=&\mathcal{G}^{B \rightarrow\ ACD}(\sigma_{ABCD}) =\max\left\{0,\frac{1}{2}\ln\frac{\cosh^4(r)}{{\rm det}\sigma_{ABCD}}\right\},
\end{eqnarray}
\begin{eqnarray}\label{S32}
\mathcal{G}^{BCD \rightarrow\ A}(\sigma_{ABCD})&=&\mathcal{G}^{ABD \rightarrow\ C}(\sigma_{ABCD})=\max\left\{0,\frac{1}{2}\ln\frac{\cosh^4(r)\cosh^{2}(2s)}{{\rm det}\sigma_{ABCD}}\right\},
\end{eqnarray}
\begin{eqnarray}\label{S34}
\mathcal{G}^{D \rightarrow\ ABC}(\sigma_{ABCD})&=&\max\left\{0,\frac{1}{2}\ln\frac{[\cosh^{2}(r)\cosh^{2}(s)+\sinh^{2}s]^{2}}{{\rm det}\sigma_{ABCD}}\right\},
\end{eqnarray}
\begin{eqnarray}\label{S35}
\mathcal{G}^{ACD \rightarrow\ B}(\sigma_{ABCD})&=&\max\left\{0,\frac{1}{2}\ln\frac{{\rm det} \sigma_{ACD}}{{\rm det}\sigma_{ABCD}}\right\},
\end{eqnarray}
where
\begin{eqnarray}\label{S313}
\nonumber \rm{det}\sigma_{ABCD}&=&\frac{1}{64}\bigg\{ 2 - 2\cosh(2r)+\cosh[2(r-s)]+6\cosh(2s)+\cosh[2(r+s)] \bigg\}^2 \\ \nonumber
\nonumber \rm{det}\sigma_{ACD}&=&\frac{1}{64}\bigg\{ 2 - 2\cosh(2r)+3\cosh[2(r-s)]+2\cosh(2s)+3\cosh[2(r+s)] \bigg\}^2.\\ \nonumber
\end{eqnarray}
These results reveal that the degree of Gaussian quadripartite steering depends not only on the squeezing parameter $r$, which governs the intrinsic quantum correlations, but also on the thermal parameter $s$ arising from Hawking radiation of the black hole. The explicit appearance of $s$ highlights the sensitivity of multipartite steering to the black hole's thermal environment near the event horizon.

\begin{figure}
\begin{minipage}[t]{0.5\linewidth}
\centering
\includegraphics[width=3.0in,height=5.3cm]{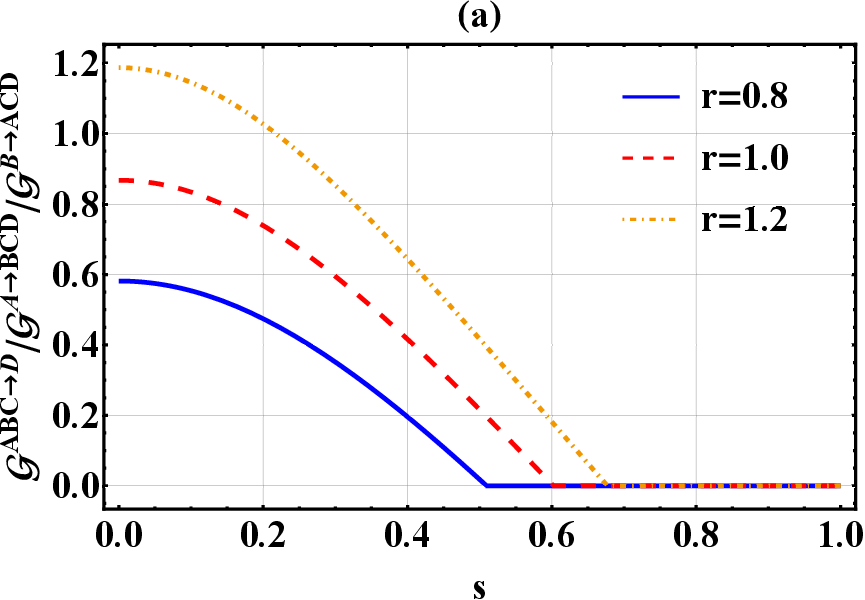}
\label{fig1a}
\end{minipage}%
\begin{minipage}[t]{0.5\linewidth}
\centering
\includegraphics[width=3.0in,height=5.3cm]{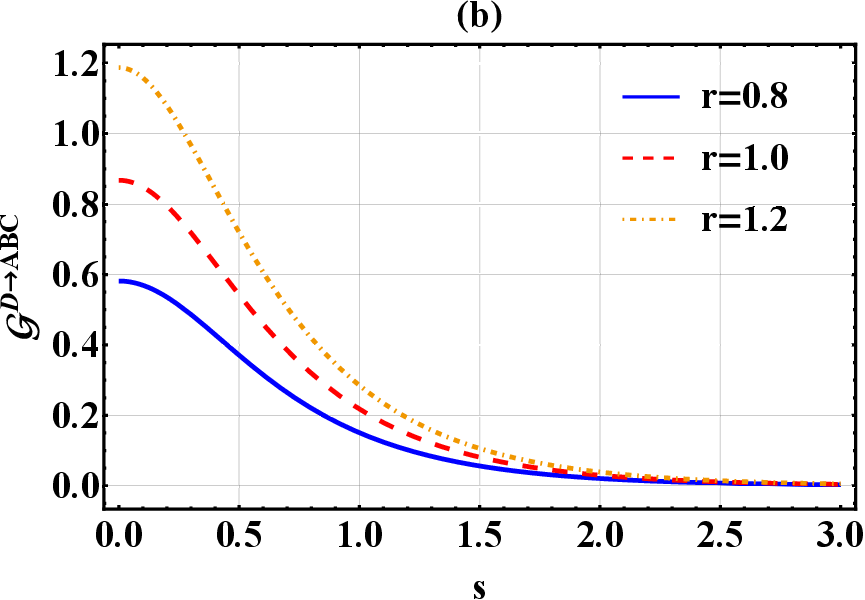}
\label{fig1c}
\end{minipage}%

\begin{minipage}[t]{0.5\linewidth}
\centering
\includegraphics[width=3.0in,height=5.3cm]{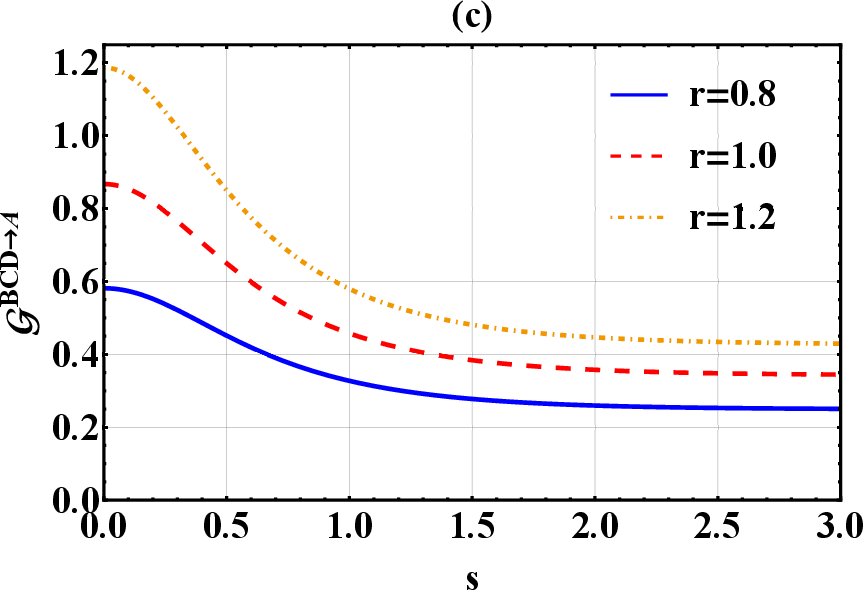}
\label{fig1a}
\end{minipage}%
\begin{minipage}[t]{0.5\linewidth}
\centering
\includegraphics[width=3.0in,height=5.3cm]{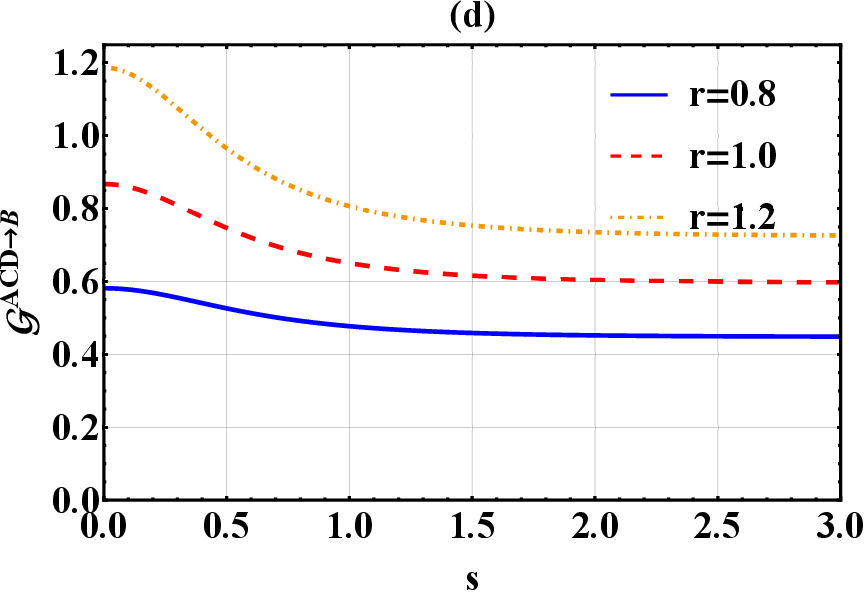}
\label{fig1c}
\end{minipage}%
\caption{Physically accessible $1\rightarrow 3$ and $3\rightarrow 1$ steering as a function of the Hawking temperature parameter $s$ for different initial parameters $r$.}
\label{F1}
\end{figure}

In Fig.\ref{F1}, we illustrate the behavior of Gaussian quadripartite steering for different $1\rightarrow 3$  and  $3\rightarrow 1$ partitions among Alice, Bob, Charlie, and David, as a function of the Hawking temperature parameter $s$ in Schwarzschild spacetime. The connection between $s$ and the Hawking temperature $T$ is given by $\cosh s=\frac{1}{\sqrt{1-e^{-\omega/T}}}$, showing that $s$ grows monotonically with $T$ for fixed $\omega$. From the plots, the results reveal several distinct features: (i) steering from non-gravitational modes to gravitational modes decreases with increasing $s$, and eventually undergoes a ``sudden death", signifying that sufficiently strong Hawking effect can entirely eliminate directional quantum correlations; (ii) steering in the opposite direction, from gravitational to non-gravitational modes, shows a smooth monotonic decline with $s$ and disappears only in the extreme black hole limit ($s\rightarrow \infty$), implying an inherent asymmetry in the influence of spacetime curvature on quantum steering; (iii) steering from hybrid partitions involving both gravitational and non-gravitational sectors toward non-gravitational mode tends to a finite asymptotic value, which depends on the initial squeezing parameter $r$. This highlights that the amount of quantum resources encoded in the initial state is decisive for the persistence of Gaussian quadripartite steering.

In contrast to quantum entanglement, quantum steering is inherently directional, a property that has been experimentally confirmed in several recent works \cite{Q13,Q14,Q15,Q16}. While Gaussian quadripartite steering is perfectly symmetric in flat spacetime, this symmetry is broken once the Hawking effect of the black hole is taken into account. As shown in Fig.\ref{F1}, we find that $\mathcal{G}^{BCD \rightarrow A}$ is consistently larger than $\mathcal{G}^{A \rightarrow BCD}$, $\mathcal{G}^{D \rightarrow ABC}$ exceeds $\mathcal{G}^{ABC \rightarrow D}$, and $\mathcal{G}^{ACD \rightarrow B}$ dominates over $\mathcal{G}^{B \rightarrow ACD}$. These trends clearly demonstrate that the curved spacetime background introduces an asymmetry into the steering dynamics. To quantitatively capture this effect, we define the Gaussian steering asymmetry between two opposite directions, such as
\begin{eqnarray}
\mathcal{G}^{\Delta}_{A|BCD}=|\mathcal{G}^{A\rightarrow BCD}-\mathcal{G}^{BCD\rightarrow A}|.
\end{eqnarray}
Analogously, the asymmetries between $\mathcal{G}^{ABC \rightarrow D}$ and $\mathcal{G}^{D \rightarrow ABC}$, as well as between $\mathcal{G}^{ACD \rightarrow B}$ and $\mathcal{G}^{B \rightarrow ACD}$, are given by
\begin{eqnarray}
\mathcal{G}^{\Delta}_{D|ABC}=|\mathcal{G}^{ABC\rightarrow D}-\mathcal{G}^{D\rightarrow ABC}|, \quad
\mathcal{G}^{\Delta}_{B|ACD}=|\mathcal{G}^{ACD\rightarrow B}-\mathcal{G}^{B\rightarrow ACD}|.
\end{eqnarray}
By analyzing these asymmetries, we can obtain a deeper understanding of the physical mechanisms that dictate the dynamics of Gaussian quadripartite steering near the event horizon.

\begin{figure}[htbp]
\centering
\includegraphics[height=1.8in,width=2.0in]{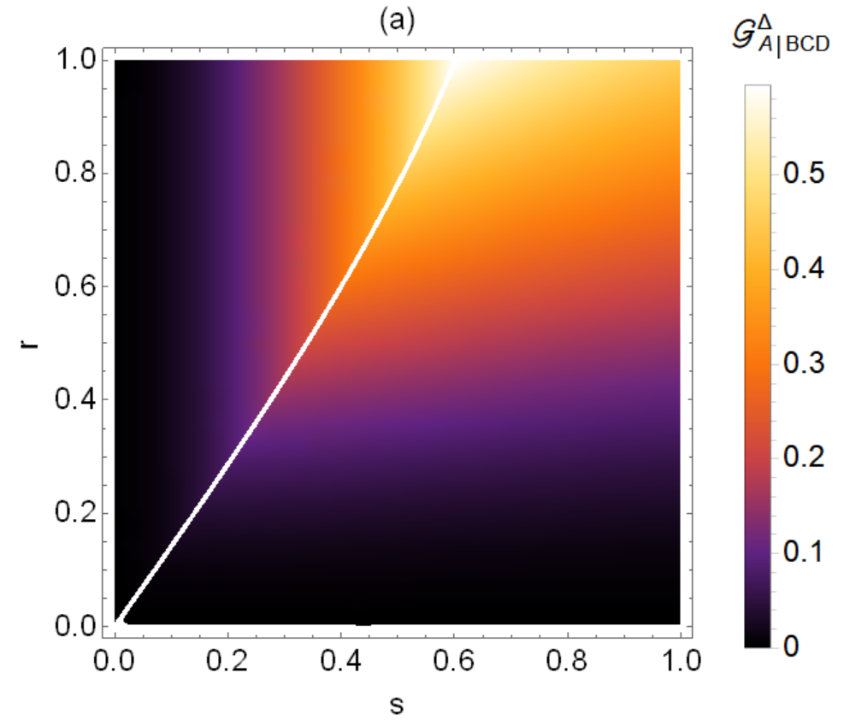}
\includegraphics[height=1.8in,width=2.0in]{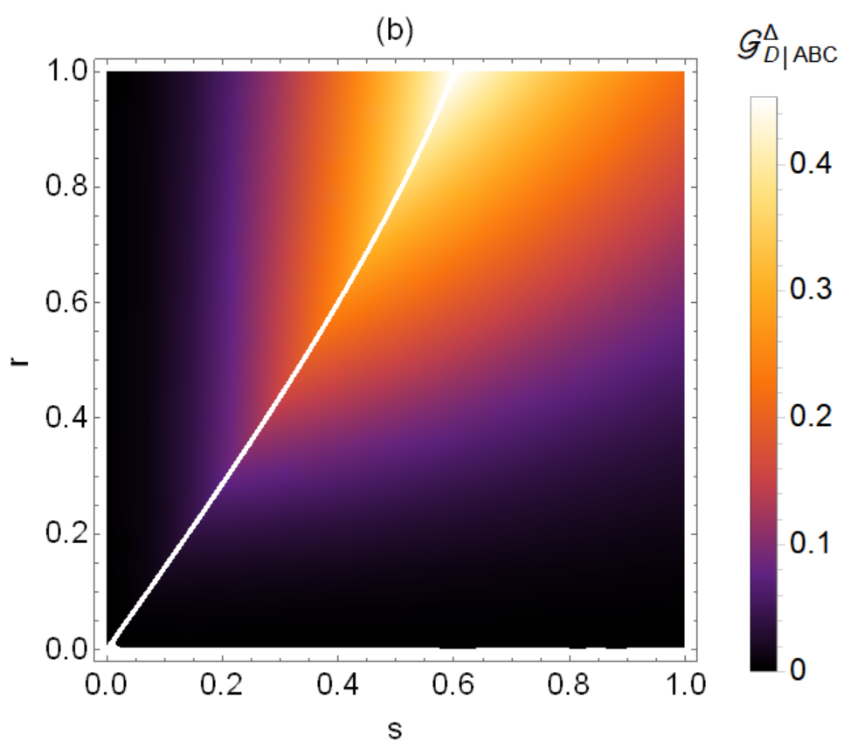}
\includegraphics[height=1.8in,width=2.0in]{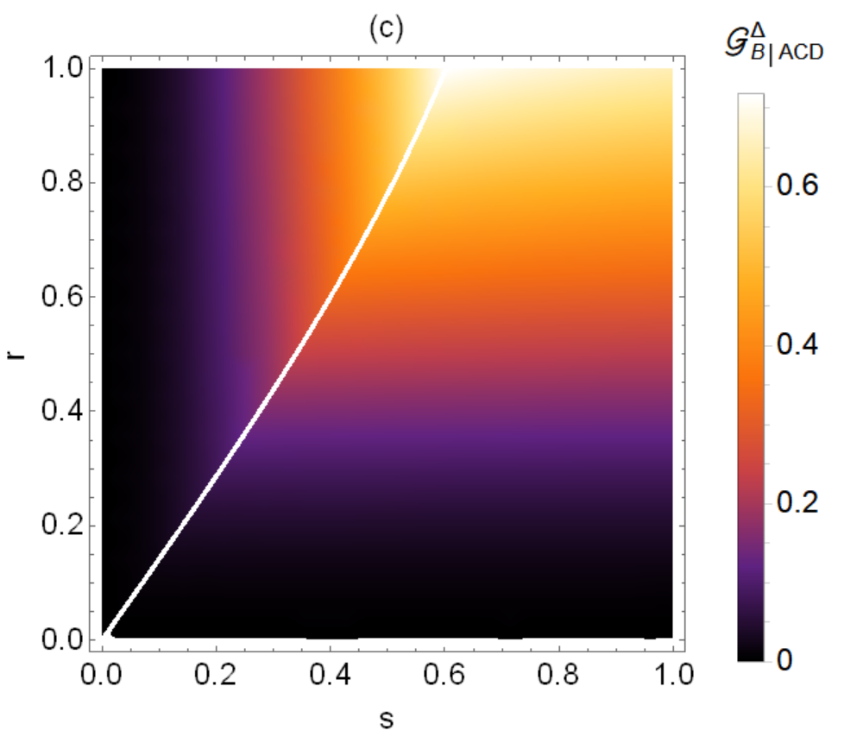}
\caption{ Gaussian steering asymmetries of accessible four-mode systems as  functions of the Hawking temperature parameter $s$ and the initial squeezing parameter $r$.
}\label{F2}
\end{figure}

In Fig.\ref{F2}, we illustrate the behavior of Gaussian steering asymmetries, namely $\mathcal{G}^{\Delta}_{A|BCD}$, $\mathcal{G}^{\Delta}_{D|ABC}$, and $\mathcal{G}^{B \rightarrow ACD}$, as functions of both the Hawking temperature parameter $s$ and the initial squeezing parameter $r$. From Eqs.(\ref{S31})-(\ref{S35}), the steering asymmetries associated with one-way steerability take the form
\begin{eqnarray}
\nonumber &&\mathcal{G}^{\Delta,1}_{A|BCD}=\frac{1}{2}\ln\frac{\cosh^4(r)\cosh^{2}(2s)}{{\rm det}\sigma_{ABCD}}, \quad \mathcal{G}^{\Delta,1}_{D|ABC}=\frac{1}{2}\ln\frac{[\cosh^{2}(r)\cosh^{2}(s)+\sinh^{2}s]^{2}}{{\rm det}\sigma_{ABCD}}, \\ \nonumber
&& \mathcal{G}^{\Delta,1}_{B|ACD}=\frac{1}{2}\ln\frac{{\rm det} \sigma_{ACD}}{{\rm det}\sigma_{ABCD}}, \nonumber
\end{eqnarray}
which monotonically decrease with $s$. In contrast, for two-way steering, the asymmetries are given by
\begin{eqnarray}
\nonumber &&\mathcal{G}^{\Delta,2}_{A|BCD}=\frac{1}{2}\ln\frac{\cosh^4(r)\cosh^{2}(2s)}{\cosh^4(r)}, \quad \mathcal{G}^{\Delta,2}_{D|ABC}=\frac{1}{2}\ln\frac{[\cosh^{2}(r)\cosh^{2}(s)+\sinh^{2}s]^{2}}{\cosh^4(r)}, \\ \nonumber
&& \mathcal{G}^{\Delta,2}_{B|ACD}=\frac{1}{2}\ln\frac{{\rm det} \sigma_{ACD}}{\cosh^4(r)}, \nonumber
\end{eqnarray}
which instead grow with $s$. The peak of steering asymmetry appears exactly at the crossover point between two-way and one-way steering in curved spacetime. Explicit evaluation shows that the optimal condition is reached at
$$s_0={\rm
arccosh}\left[ 2 \sqrt{\frac{\cosh^{2}(r)}{3 + \cosh(2r)}} \right],
$$
which corresponds to the sudden vanishing of steering from non-gravitational to gravitational modes as $s$ increases. This behavior signals the onset of a transition from bidirectional to unidirectional steerability. In other words, the attainment of maximal steering asymmetry serves as a clear indicator of the crossover between one-way and two-way steerability of the four-mode Gaussian state under the influence of Hawking radiation. Fig.\ref{F2} further reveals that $s_0$, the location of maximal asymmetry, shifts to larger values with increasing squeezing parameter $r$. This indicates that the initial quantum resources encoded in the squeezing play a decisive role in shaping the asymmetry of Gaussian steering under relativistic effects. In particular, the interplay between Hawking-induced thermal noise and state preparation strongly influences the redistribution of quantum correlations, underscoring how both the Hawking temperature and the squeezing determine steering behavior in curved spacetime.

\subsection{Generation of Gaussian quadripartite steering between initially uncorrelated modes }
In this subsection, we investigate how Gaussian quadripartite steering can arise among initially uncorrelated modes in the Schwarzschild black hole. When mode $D$ is traced out, the covariance matrix describing the joint system of Alice, Bob, Charlie, and anti-David takes the form
\begin{eqnarray}\label{B3}
\sigma_{ABC\bar D}(r,s) &=&
\left(\begin{array}{cccc}
\mathcal{\sigma}_{A}(r) & \mathcal{E}_{AB}(r) & \mathcal{E}_{AC}(r) & \mathcal{E}_{A\bar D}(r,s) \\
\mathcal{E}_{AB}^{\sf T}(r) &  \mathcal{\sigma}_{B}(r) & \mathcal{E}_{BC}(r) & \mathcal{E}_{B\bar D}(r,s) \\
\mathcal{E}^{\sf T}_{AC}(r) & \mathcal{E}_{BC}^{\sf T}(r) & \mathcal{\sigma}_{C} (r)& \mathcal{E}_{C\bar D}(r,s) \\
\mathcal{E}^{\sf T}_{A\bar D}(r,s) & \mathcal{E}^{\sf T}_{C\bar D}(r,s) & \mathcal{E}^{\sf T}_{C\bar D}(r,s) &  \mathcal{\sigma}_{\bar D}(r,s)
\end{array}
\right)\,.
\end{eqnarray}
By applying Eqs.(\ref{wA1}) and (\ref{wA2}), the corresponding steering quantifiers among $A$, $B$, $C$, and $\bar D$ are obtained as
\begin{eqnarray}\label{S39}
\mathcal{G}^{ABC \rightarrow\ \bar D}(\sigma_{ABC\bar{D}})&=&\mathcal{G}^{C \rightarrow\ AB\bar{D}}(\sigma_{ABC\bar{D}})=\mathcal{G}^{A \rightarrow\ BC\bar{D}}(\sigma_{ABC\bar{D}})\\
&=&\mathcal{G}^{B \rightarrow\ \nonumber AC\bar{D}}(\sigma_{ABC\bar{D}})=\max\left\{0,\frac{1}{2}\ln\frac{\cosh^{4}(r)}{\rm{det}\sigma_{ABC\bar{D}}}\right\}=0,
\\\mathcal{G}^{\bar D \rightarrow\ ABC}(\sigma_{ABC\bar{D}})&=&\max\left\{0,\frac{1}{2}\ln\frac{[\cosh^{2}(s)+\cosh^{2}(r)\sinh^{2}{s}]^{2}}{\rm{det}\sigma_{ABC\bar{D}}}\right\}=0,
\\\mathcal{G}^{AB\bar{D} \rightarrow\ C}(\sigma_{ABC\bar{D}})&=&\mathcal{G}^{BC\bar D \rightarrow\ A}(\sigma_{ABC\bar{D}})=\max\left\{0,\frac{1}{2}\ln\frac{\cosh^{4}(r)\cosh^{2}(2s)}{\rm{det}\sigma_{ABC\bar{D}}}\right\},
\\\mathcal{G}^{AC\bar{D} \rightarrow\ B}(\sigma_{ABC\bar{D}})&=&\max\left\{0,\frac{1}{2}\ln\frac{\rm{det}\sigma_{AC\bar{D}}}{\rm{det}\sigma_{ABC\bar{D}}}\right\},
\end{eqnarray}
where
\begin{eqnarray}\label{S313}
\nonumber \rm{det}\sigma_{ABC\bar D}&=&\frac{1}{64}\bigg\{ -2 + 2\cosh(2r)+\cosh[2(r-s)]+6\cosh(2s)+\cosh[2(r+s)] \bigg\}^2,\\ \nonumber
\rm{det}\sigma_{AC\bar D}&=&\frac{1}{64}\bigg\{ -2 + 2\cosh(2r)+3\cosh[2(r-s)]+2\cosh(2s)+3\cosh[2(r+s)] \bigg\}^2.
\end{eqnarray}

Next, by tracing out mode $C$, the covariance matrix of the system composed of Alice, Bob, David, and anti-David can be written as
\begin{eqnarray}\label{B33}
\sigma_{ABD\bar D}(r,s) &=&
\left(\begin{array}{ccccc}
\mathcal{\sigma}_{A}(r) & \mathcal{E}_{AB}(r) & \mathcal{E}_{AC}(r) & \mathcal{E}_{A\bar D}(r,s) \\
\mathcal{E}^{\sf T}_{AB}(r) & \mathcal{\sigma}_{B}(r) & \mathcal{E}_{BC}(r) & \mathcal{E}_{B\bar D}(r,s) \\
\mathcal{E}^{\sf T}_{AD}(r) & \mathcal{E}^{\sf T}_{BD}(r) &  \mathcal{\sigma}_{D}(r) &\mathcal{E}_{D\bar D}(r,s) \\
\mathcal{E}^{\sf T}_{A\bar D}(r,s) & \mathcal{E}^{\sf T}_{B\bar D}(r,s) & \mathcal{E}^{\sf T}_{D\bar D}(r,s) &\mathcal{\sigma}_{\bar D}(r,s)
\end{array}
\right)\,.
\end{eqnarray}
The corresponding analytic expressions of Gaussian steering for this configuration are obtained as
\begin{eqnarray}\label{SS39}
\mathcal{G}^{ABD \rightarrow\ \bar D}(\sigma_{ABD\bar{D}})&=&\mathcal{G}^{AB\bar{D} \rightarrow\ D}(\sigma_{ABD\bar{D}})=\max\left\{0,\frac{1}{2}\ln\cosh^{2}(2s) \right\},
\\\mathcal{G}^{\bar D \rightarrow\ ABD}(\sigma_{ABD\bar{D}})&=&\max\left\{0,\frac{1}{2}\ln\frac{[\cosh^{2}(s)+\cosh^{2}(r)\sinh^{2}(s)]^{2}}{\cosh^{4}(r)}\right\},
\\\mathcal{G}^{D \rightarrow\ AB\bar{D}}(\sigma_{ABD\bar{D}})&=&\max\left\{0,\frac{1}{2}\ln\frac{[\cosh^{2}(r)\cosh^{2}(s)+\sinh^{2}(s)]^{2}}{\cosh^{4}(r)}\right\},
\\\mathcal{G}^{BD\bar D \rightarrow\ A}(\sigma_{ABC\bar{D}})&=&\max\left\{0,\frac{1}{2}\ln\frac{\cosh^{2}(2r)}{\cosh^{4}(r)}\right\},
\\\mathcal{G}^{AD\bar{D} \rightarrow\ B}(\sigma_{ABD\bar{D}})&=&\mathcal{G}^{A \rightarrow\ BD\bar{D}}(\sigma_{ABD\bar{D}})=\mathcal{G}^{B \rightarrow\ AD\bar{D}}(\sigma_{ABC\bar{D}}) \\ \nonumber
&=&\max\left\{0,\frac{1}{2}\ln\frac{\cosh^{4}(r)}{\cosh^{4}(r)}\right\}=0.
\end{eqnarray}
From the above equation, we find that $\mathcal{G}^{ABD \rightarrow\ \bar D}(\sigma_{ABD\bar{D}})$ and $\mathcal{G}^{AB\bar{D} \rightarrow\ D}(\sigma_{ABD\bar{D}})$
are governed exclusively by the Hawking temperature parameter $s$, remaining independent of the initial squeezing parameter $r$. Conversely, $\mathcal{G}^{BD\bar D \rightarrow\ A}(\sigma_{ABC\bar{D}})$ depends solely on the initial squeezing parameter $r$, being completely insensitive to the thermal effects characterized by Hawking radiation. This asymmetry highlights the distinct roles played by gravitationally induced Hawking radiation and initial quantum correlations in shaping the directional distribution of Gaussian steering among multipartite modes.

\begin{figure}
\begin{minipage}[t]{0.5\linewidth}
\centering
\includegraphics[width=3.0in,height=5.3cm]{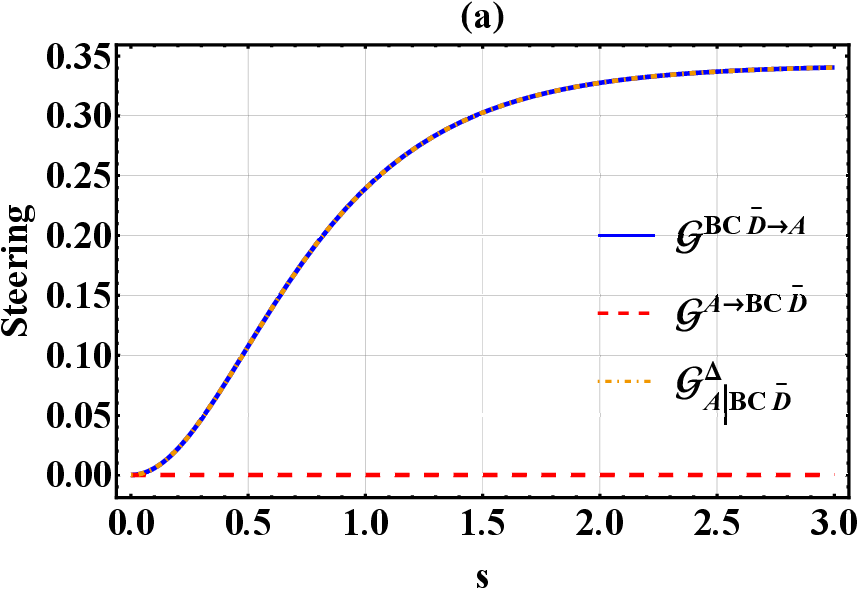}
\label{fig1a}
\end{minipage}%
\begin{minipage}[t]{0.5\linewidth}
\centering
\includegraphics[width=3.0in,height=5.3cm]{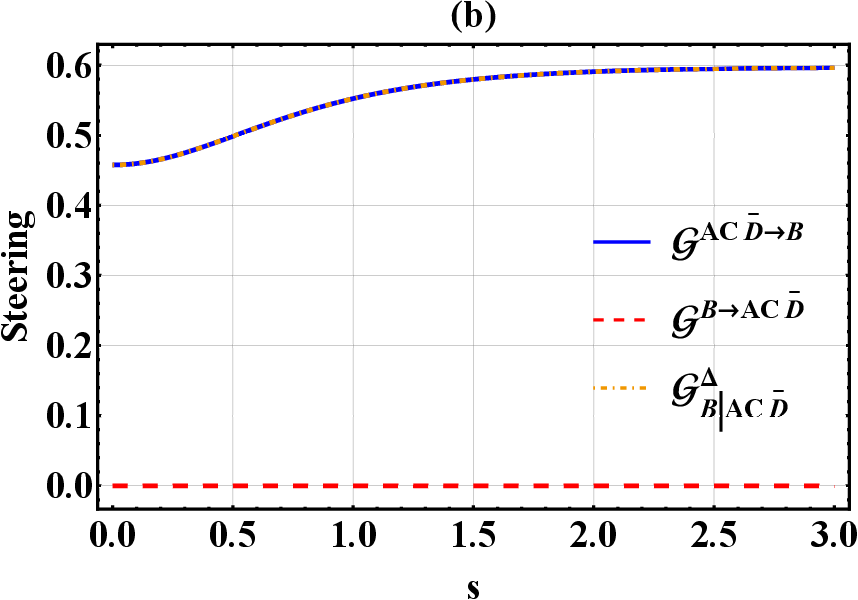}
\label{fig1c}
\end{minipage}%

\begin{minipage}[t]{0.5\linewidth}
\centering
\includegraphics[width=3.0in,height=5.3cm]{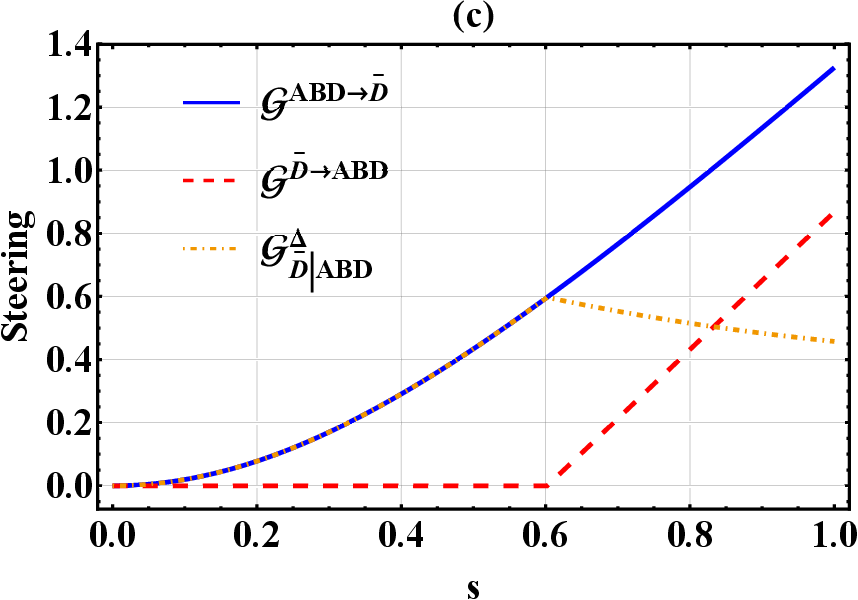}
\label{fig1a}
\end{minipage}%
\begin{minipage}[t]{0.5\linewidth}
\centering
\includegraphics[width=3.0in,height=5.3cm]{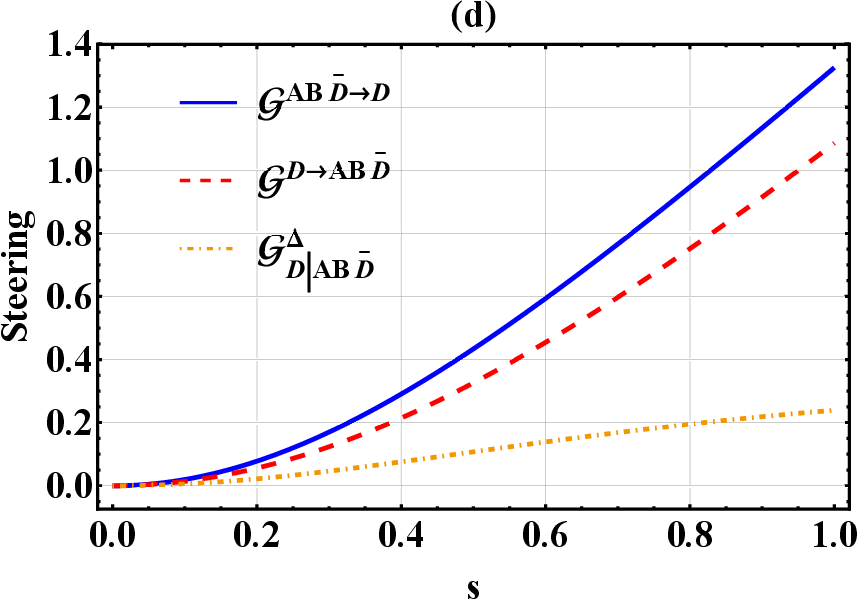}
\label{fig1c}
\end{minipage}%
\caption{Physically inaccessible steering and steering asymmetry as a function of the Hawking temperature parameter $s$ for a fixed initial parameter $r=1$. }
\label{F3}
\end{figure}

In Fig.\ref{F3}, we display the Gaussian steering and the associated asymmetry of physically inaccessible systems as a function of the Hawking temperature parameter $s$, with the initial squeezing fixed at $r=1$. It is evident from the figure that the steering quantity $\mathcal{G}^{AC\bar{D} \rightarrow\ B}(\sigma_{ABC\bar{D}})$ grows monotonically from its initial value $\ln[\cosh(2r)\rm{sech}^2(r)]$. This growth stems from the finite contribution of the accessible steering $\mathcal{G}^{AC \rightarrow\ B}(\sigma_{ABC})$ present at $s=0$. By contrast, the remaining inaccessible steering measures vanish at the beginning and only emerge progressively as $s$ increases. Notably, $\mathcal{G}^{\bar{D} \rightarrow\ ABD}(\sigma_{ABC\bar{D}})$ displays a sudden onset with rising $s$, where the steering asymmetry reaches its peak, signifying the crossover from unidirectional  to bidirectional  steerability. Furthermore, all inaccessible steering  induced by the Hawking effect are inherently asymmetric, with their behavior depending crucially on the steering direction.
Compared with multipartite quantum entanglement and coherence, the redistribution of multipartite quantum steering in the Schwarzschild black hole background exhibits a stronger directional dependence and reveals richer and more profound physical characteristics.

\section{ CONCLUSIONS  \label{GSCDGE}}
In this work, we have investigated the distribution of Gaussian quadripartite steering in the Schwarzschild black hole background. Our model involves five modes: modes $A$, $B$, and $C$, associated with Kruskal observers Alice, Bob, and Charlie, respectively; mode $D$, corresponding to the Schwarzschild observer David located outside the event horizon; and mode $\bar{D}$, monitored by anti-David inside the event horizon. By employing the Bogoliubov transformation linking Kruskal and Schwarzschild vacua, we obtain the phase-space dynamics of an initially Gaussian four-mode state under the action of Hawking radiation. For the physically accessible system, several key behaviors emerge. \textbf{(i) Sudden death and maximal asymmetry:} steering from non-gravitational observers to gravitational ones is rapidly suppressed as the Hawking parameter increases, eventually undergoing a sudden death. This disappearance coincides with the point of maximal steering asymmetry, which serves as a clear marker of the crossover from two-way to one-way steerability. \textbf{(ii) Directional decay of gravitational to non-gravitational steering:} in the reverse direction, steering from gravitational to non-gravitational modes decreases monotonically and vanishes only in the extremal black hole limit, reflecting the intrinsic directionality imposed by spacetime curvature. \textbf{(iii) Persistence of hybrid steering:}
steering involving hybrid partitions, where gravitational and non-gravitational modes jointly steer a non-gravitational subsystem, approaches a finite asymptotic value determined by the initial squeezing parameter. This demonstrates that the persistence of Gaussian quadripartite steering crucially depends on the quantum resources encoded in the initial state.

For the physically inaccessible subsystems, the Hawking effect of the black hole can induce inaccessible steering that displays a pronounced and inherently direction-dependent asymmetry. Remarkably, this steering may experience a "sudden birth" phenomenon, during which the asymmetry reaches its maximum, signaling a transition from one-way to two-way steerability. From a broader perspective, this behavior reflects the fact that the Hawking effect does not simply degrade quantum correlations but actively redistributes multipartite steering among accessible and inaccessible sectors. Such redistribution reshapes the balance between gravitationally influenced and unaffected modes, thereby revealing a fundamental mechanism by which black holes govern the flow and accessibility of quantum information across event horizons.

\begin{acknowledgments}
This work is supported by the National Natural
Science Foundation of China (12575056), the Young Elite Scientist Sponsorship Program by Guizhou Science and Technology Association (Grant No. GASTYESS202424), the Discipline-Team of
Liupanshui Normal University of China (Grant No. LPSSY2023XKTD11), and the Special Fund for Basic Scientific Research of Provincial Universities in Liaoning under grant NO.LS2024Q002.
\end{acknowledgments}

\end{document}